# Raman-assisted crystallography reveals end-on peroxide intermediates in a non-heme iron enzyme.

Gergely Katona[a], Philippe Carpentier[a], Vincent Nivière[c], Patricia Amara[a], Virgile Adam[b], Jérémy Ohana[a], Nikolay Tsanov[a], & Dominique Bourgeois[a,b]

[a]*IBS, Institut de Biologie Structurale Jean-Pierre Ebel; CEA; CNRS; Université Joseph Fourier, 41 rue Jules Horowitz, F-38027 Grenoble, France*

[b]*European Synchrotron Radiation Facility, 6 rue Jules Horowitz, BP 220, 38043 Grenoble Cedex, France*

[c]*Laboratoire de Chimie et Biologie des Métaux, iRTSV-CEA/CNRS/Université J. Fourier, UMR 5249, 17 rue des Martyrs, 38054 Grenoble Cedex 9, France*

*Correspondence and request for materials should be addressed to D. B. (e-mail dominique.bourgeois@ibs.fr).*

**One-sentence summary:** Iron(III)-(hydro)peroxo transient species in superoxide reductase adopt end-on configurations that promote Fe-O bond cleavage during catalysis.

**Iron-peroxide intermediates are central in the reaction cycle of many iron containing biomolecules. We trapped iron(III)-(hydro)peroxo species in crystals of superoxide reductase (SOR), a non-heme mononuclear iron-enzyme that scavenges superoxide radicals. X-ray diffraction data at 1.95-angström resolution and Raman spectra recorded *in crystallo* revealed iron-(hydro)peroxo intermediates with the (hydro)peroxo group bound end-on. The dynamic SOR active site promotes the formation of transient hydrogen bond networks, which presumably assist the cleavage of the Fe-O bond in order to release the reaction product, hydrogen peroxide.**

The interaction of dioxygen with iron containing proteins is important in many biological processes, including transport, metabolism, respiration and cell protection. The reaction of oxygen or its reduced derivatives, superoxide and hydrogen peroxide, with iron-enzymes often involves short-lived iron-peroxide intermediates along the reaction cycle (*1, 2*). Heme-based peroxidases, catalases, and many oxygenases promote heterolytic cleavage of the peroxide oxygen-oxygen bond to form high-valence reactive iron-oxo species. In contrast, other iron-enzymes such as superoxide reductase (SOR) (*3, 4*) are fine-tuned to cleave the iron-oxygen bond and avoid the formation of potentially harmful iron-oxo species. Although the protein, the metal configuration, and the solvent environment have been shown to play a role, the mechanisms by which iron-peroxide intermediates are processed are not fully understood (*1, 2*). Despite pioneering studies on heme-proteins (*5-7*), structural data revealing peroxide species in non-heme mononuclear iron-enzymes have remained scarce (*8*). Here, we have developed a new approach in which kinetic crystallography (*9*) was assisted by "*in crystallo*" Raman spectroscopy (*10*) to characterize (hydro)peroxo species in superoxide reductase. SOR is found in some air-sensitive bacteria or archaea and converts the toxic superoxide anion radical ($O_2^{\bullet-}$) into hydrogen peroxide ($H_2O_2$) via a one-electron reduction pathway: $O_2^{\bullet-} + 2H^+ + SOR(Fe^{2+}) \rightarrow H_2O_2 + SOR(Fe^{3+})$ (*3, 4, 11*). The SOR catalytic domain displays an immunoglobulin-like fold (*12, 13*) encompassing an iron atom coordinated to four equatorial histidines and one axial cysteine, thus bearing structural resemblance with the ubiquitous cytochromes P450s. However, contrary to P450s, the ferrous enzyme is stable under atmospheric conditions, with a vacant, solvent-exposed, sixth coordination site where $O_2^{\bullet-}$ is thought to bind (*14*). Investigations of various SOR adducts (*12, 14, 15*), pulse-radiolysis studies (*16-18*), and resonance Raman spectroscopy (*19, 20*) have suggested an inner-sphere catalytic mechanism involving the formation of transient iron(III)-(hydro)peroxo species. As described for similar enzymes (*2, 21*), protonation steps play a crucial role in governing the chemistry that occurs at the SOR active site (*11, 16, 17*). In SOR from the sulfate reducing-bacterium *Desulfoarculus baarsii* (*4, 12*), a first iron(III)-peroxo intermediate has been proposed to be rapidly protonated (in ~100 μs), forming a more stable iron(III)-hydroperoxo species (*17*). A second protonation then occurs, possibly promoted by a water molecule (*22*) and yields the $H_2O_2$ product through a dissociative mechanism in which Glu$^{47}$ ultimately binds to the oxidized enzyme (*13, 17*). Thus, SOR avoids heterolytic cleavage of the O-O bond, preventing the formation of oxo-ferryl compounds. To date, the structure of the iron-peroxide species that can be accommodated within the SOR active site and the mechanism governing the decisive second protonation step have remained elusive (*11*). The structural data described below clearly reveal a series of end-on iron(III)-(hydro)peroxo species involved in tight hydrogen bond networks, and allow us to propose a mechanism for proton assisted release of $H_2O_2$ in SOR.

Mononuclear iron-peroxide complexes are generally obtained by reacting iron(II) with excess $H_2O_2$ (*23*). However, to minimize Fenton-driven generation of toxic hydroxyl radicals, crystalline SOR was first oxidized with hexachloroiridate(IV), and then exposed to $H_2O_2$ for 3 minutes before freezing (*24*). As the isolation of iron(III)-peroxide complexes is hampered by their high reactivity, crystallographic data were collected using the mutant enzyme E114A, in which, as described for the E47A variant (*19, 20*), these intermediates are stabilized (**supporting online text**). Comparison of the native crystal structures of the wild type and mutant enzymes revealed that the loss of the E114 side chain does not alter the overall enzyme structure (**Table 1**).

The asymmetric unit in SOR-E114A crystals contains four monomers denoted A to D. Upon soaking with $H_2O_2$, diffraction data to 1.95 Å resolution (*24*) (**Table S1**) revealed clear elongated features in the electron density maps, that are consistent with the formation of end-on ($\eta^1$) peroxide species in monomers B, C and D (**Figure 1 & 2**), while monomer A did not react significantly. To verify the chemical nature of the observed species, we developed a Raman spectrometer to analyze cryo-cooled crystals under conditions identical to those used for X-ray data collection (*24, 25*) **(Figure S1)**. Upon $H_2O_2$ treatment, two $^{18}O$ isotope-sensitive main bands at ~567 cm$^{-1}$ and ~838 cm$^{-1}$ appeared in the Raman spectra of SOR crystals (**Figure 3**). Although these bands probably involve the coupling of a number of vibrational modes, they fall within the expected range for ν(Fe-OO(H)) and ν(O-O) frequencies of iron-peroxide species, respectively (*26*, **supporting online text**). Importantly, they are not specific to the crystalline phase, as they also appeared using solution samples similarly treated with hydrogen peroxide (**supporting online text**). In addition, Raman spectra from crystals *exposed to X-rays* (*27*) showed the same signature as unexposed crystals, ruling out the possibility of significant photo-alteration during data collection (it is known that the solvent-exposed SOR active site is sensitive to reduction by X-ray induced photo-electrons (*12*)). Overall, *in crystallo* Raman spectra strongly suggested the build-up of iron-peroxide species in the crystal. To assess the protonation state of these species, we performed Density Functional Theory (DFT) calculations (**supporting online text**) on model SOR active sites based on the X-ray structures determined in this work. In monomers B and D, these calculations clearly favor high-spin $\eta^1$ hydroperoxo species that are protonated at the distal oxygen, consistent with pulse-radiolysis studies that suggested rapid protonation of the SOR iron-peroxo species even at the basic pH (pH = 9) used in our work (*17*). In monomer C, an $\eta^1$ species is also favored but its protonation state cannot be firmly established.

Final X-ray models of SOR monomers B, C and D show end-on iron(III)-peroxide species in three different configurations that all display the distal oxygen pointing towards His$^{119}$ to accommodate steric constraints imposed by the protein matrix. Thus, the atoms $C_{\beta,Cys116}$-$S_{\gamma,Cys116}$-Fe-O-O are non co-planar, resulting in non-optimal π orbital overlap and contributing to weaken the iron-oxygen bond, as also suggested by the long Fe-O distances found in our DFT calculations (**Table 1**).

In SOR, a solvent exposed flexible loop (residues 45 to 49, called "LID" thereafter) is located near the active site and includes Lys$^{48}$, an evolutionary conserved residue critical for efficient catalysis (*16, 17*). The formation of iron-peroxide species modified the conformation of Lys$^{48}$ relative to the structure of SOR in the native reduced state (**supporting online text**). In monomer C, the hydroperoxo moiety only interacts with the active iron and the LID loop displays a "locked open" conformation possibly due to weak crystal lattice contacts (**supporting online text**). This conformation of the LID prevents Lys$^{48}$ from interacting with the hydroperoxo moiety, leaving the side chain of this residue in a disordered state. In contrast, in monomer B, the LID loop is found "locked closed" and Lys$^{48}$ facilitates a tight hydrogen bond network around the distal oxygen of the peroxide moiety that also includes two water molecules (Wat$^{10}$ and Wat$^{11}$) (**Figure 2**). The positively charged amino group of Lys$^{48}$ (**supporting online text**) attracts the peroxide ligand, presumably inducing a stretch of the S$_{\gamma,Cys116}$-Fe-O-O moiety that may further weaken the Fe-O bond. In monomer D, the side chain of Lys$^{48}$ slightly rotates away from the hydroperoxo moiety, and the two water molecules Wat$^{10}$ and Wat$^{11}$ are still observed, together with a third molecule (Wat$^{12}$) that may play a stabilizing role. However, Wat$^{10}$ now forms a hydrogen bond with the amino group of Ala$^{45}$, while Wat$^{11}$ moves slightly towards the iron so that it interacts with both the proximal and distal oxygen atoms of the hydroperoxo moiety. Wat$^{11}$ is therefore in a favorable position to donate a proton to the proximal oxygen atom, allowing the formation and release of hydrogen peroxide. This is a crucial step that differentiates SOR from heme enzymes where protonation occurs at the distal oxygen, liberating water and oxo-ferryl species (*2*). Simultaneously, a combination of subtle rearrangements of the iron-coordinating histidines shifts the iron into the plane defined by the four equatorial coordinating nitrogens (**Table 1**) (*28*). This possibly facilitates access of Wat$^{11}$ to the metal and the proximal oxygen.

Our data highlight the dynamic behavior of the SOR active site en-route to product formation (**Figure 4, Movie S1**). Monomer C may be viewed as an early state along the reaction coordinate, that precedes the conformational rearrangements leading to the protonation of the HOO$^-$ adduct. We suggest that this state is stabilized in the crystal due to the "locked" configuration of the LID loop. In contrast, monomers B and D reveal subsequent activated configurations, emphasizing the catalytic role of Lys$^{48}$. The observation of Wat$^{11}$ in the immediate environment of the hydroperoxo species in these monomers strongly supports the hypothesis that this water molecule is the proton donor for product formation and release. We propose that Lys$^{48}$ hydrogen bonds to Wat$^{11}$ and imports it into the SOR active site in a motion promoted by electrostatic attraction of the positively charged amino group to the hydroperoxo ligand. Once anchored in the vicinity of the proximal oxygen, Wat$^{11}$ becomes more acidic due to the interaction with the amino group of Lys$^{48}$. Protonation of the proximal oxygen is probably simultaneous with the cleavage of the iron-oxygen bond, and Wat$^{11}$ may immediately replace the hydrogen peroxide product in the form of a hydroxide ion, until Glu$^{47}$ binds to the iron, as already suggested (*18, 22*).

Superoxide reductase illustrates the key role played by subtle protein motions in enzyme catalysis (*29*). In crystalline SOR, the flexible LID loop adopts various conformations, suggesting that there is little free energy difference between disordered (entropy-driven) states for this loop, and ordered (enthalpy-driven) ones where Lys$^{48}$ is stabilized by transient H-bonding networks. Our data are consistent with the idea of a breathing of the LID loop that serves to import catalytically competent water molecules into the SOR active site (*29*). In the crystal, local packing forces may slightly modify the thermodynamic energy balance, selecting different conformations in each monomer.

The structural observations described in this work are obviously not sufficient to entirely account for the specific reactivity of SOR towards breakage of the iron-oxygen bond. Finely tuned electron donation by the trans thiolate ligand (Cys$^{116}$) is expected to precisely adjust the strength of this bond (*11*). Furthermore, several lines of evidence indicate that the iron spin state greatly modulates the strength of the iron-oxygen and oxygen-oxygen bonds in iron(III)-peroxide complexes (*1, 11, 21, 23*)**.** Whereas many heme catalysts that promote cleavage of the O-O bond involve low-spin states of the iron, SOR (*19, 20, 30*) and the oxygen carrier di-iron hemerythrin (*31*) are unique in that they involve a high-spin (S=5/2) iron state (*30*, **supporting online text**). Interestingly, SOR and hemerythrin share structural and spectroscopic properties: in oxy-hemerythrin an end-on iron-peroxide species stabilized by a strong hydrogen bond is also observed (*31*). In addition, the Raman vibrations measured for SOR and oxy-hemerythrin are relatively similar and imply a weaker Fe-O bond and a stronger O-O bond when compared to low-spin iron-peroxide model compounds known to favor heterolytic cleavage of the O-O bond (*26*).

In conclusion, we have observed three transient peroxide species in superoxide reductase that display end-on configurations. The data suggest a possible mechanism for hydrogen peroxide formation, highlighting the role of a key water molecule finely controlled by the enzyme dynamics. The revealed conformational transitions provide a strong basis for further computational and structural investigations of the mechanism of superoxide scavenging by SOR and may facilitate the design of bio-mimetic catalysts.

**Table 1 Geometry of the active site**

|  | WT-SOR | E114A-SOR reduced | E114A-SOR peroxide intermediates | DFT calculation |
|---|---|---|---|---|
| **Monomer A** | | | | |
| Fe-S (Å) | 2.4 | 2.4 | 2.5 | |
| Fe from His plane* | 0.4 | 0.4 | 0.3 | |
| **Monomer B** | | | | |
| Fe-S (Å) | 2.4 | 2.4 | 2.5 | 2.48 |
| Fe-O1 (Å) | | | 2.0 | 2.19 |
| Fe-O1-O2 (°) | | | 126 | 125 |
| Cβ-S-O1-O2 (°) | | | 140 | 168 |
| Fe from His plane* | 0.4 | 0.4 | 0.3 | 0.10 |
| **Monomer C** | | | | |
| Fe-S (Å) | 2.4 | 2.5 | 2.5 | 2.44 |
| Fe-O1 (Å) | | | 2.0 | 1.94 |
| Fe-O1-O2 (°) | | | 126 | 123 |
| Cβ-S-O1-O2 (°) | | | 132 | 114 |
| Fe from His plane* | 0.5 | 0.4 | 0.3 | 0.16 |
| **Monomer D** | | | | |
| Fe-S (Å) | 2.4 | 2.5 | 2.6 | 2.49 |
| Fe-O1 (Å) | | | 2.0 | 2.22 |
| Fe-O1-O2 (°) | | | 122 | 123 |
| Cβ-S-O1-O2 (°) | | | 112 | 99 |
| Fe from His plane* | 0.4 | 0.3 | 0.0 | 0.11 |

* Distance from the plane defined by the coordinating N atoms of the equatorial histidines in Ångströms. Increasing value indicates an iron position closer to Cys[116].

**Figure legends.**

**Figure 1:** Structural overview of superoxide reductase. The X-ray structure of the SOR-E114A homodimer in the native reduced state is shown as *magenta* (monomer A) and *cyan* (monomer B) ribbons with the exception of the LID loop (residues 45-49), which is colored in *dark green* and *orange* for monomer A and B, respectively. Reduced and oxidized iron atoms are shown as *green* and *orange* balls, respectively. The active site of monomer B upon addition of $H_2O_2$ is highlighted in the inset. The residues coordinating the active iron ($His^{49}$, $His^{69}$, $His^{75}$, $His^{119}$ and $Cys^{116}$) as well as $Lys^{48}$ are represented as sticks. The bound peroxide ligand is shown as a *red* stick. Water molecules are shown as *red* balls. In order to support the diatomic nature of the peroxide intermediate, simulated annealed $F_{obs}$-$F_{calc}$ maps omitting the distal or proximal oxygens of the O-O moiety, respectively, were calculated. The two maps are displayed in *green* (distal) and *orange* (proximal) at a contour level of 3.0 σ.

**Figure 2:** Structure of the SOR-peroxide intermediates. Stereo views of the peroxide-bound SOR active sites in monomers C, B and D are shown in (**A**), (**B**) and (**C**), respectively. Final $2F_{obs}$-$F_{calc}$ maps (*blue*, contoured at 1.0 σ), simulated annealed $F_{obs}$-$F_{calc}$ maps omitting the peroxo moiety and associated water molecules (*green*, contoured at 4.5 σ), and simulated annealed $F_{obs}$-$F_{calc}$ maps omitting only $Lys^{48}$ (*orange*, contoured at 3.5 σ) are shown, overlaid on the refined models of the SOR-peroxide intermediates. Hydrogen bonds and iron coordination are shown as *blue* and *black* dashed lines, respectively.

**Figure 3:** Raman spectra of SOR crystals. After reaction with $H_2O_2$, the E114A-SOR mutant reveals bands at ~567 cm$^{-1}$ and ~838 cm$^{-1}$, which are isotopically shifted to ~563 cm$^{-1}$ and ~802 cm$^{-1}$ in the presence of $H_2^{18}O_2$ (vertical grey lines). Similar Raman bands and $^{18}O$ isotopic shifts are observed in solution experiments (**Figure S2**). E114A-SOR in the native reduced form does not exhibit these bands, neither do crystals oxidized by

hexachloroiridate(IV). The peaks at ~567 cm$^{-1}$ and ~838 cm$^{-1}$ are not significantly affected by exposure to an X-ray dose of $3 \times 10^5$ Gy, that is approximately the same dose as used for data collection.

**Figure 4:** <u>Proposed mechanism of superoxide reduction by SOR</u>. The proposed catalytic cycle begins with the reduced pentacoordinated active site (**1**). Superoxide binds (**2**) and gets reduced (*red arrow*) to an unprotonated peroxo species (**3**). Previous data suggest that this intermediate could correspond to a high-spin side-on peroxo-Fe$^{3+}$ species (*19, 20*, **supporting online text**). The first protonation step leads to the configuration observed in monomer C (**4**). Lys$^{48}$ and two water molecules are recruited (**5**) yielding the configuration observed in monomer B, that subsequently rearranges to give the configuration of monomer D (**6**). At this point protonation of the proximal oxygen atom (*red arrow*) is facilitated by the key water molecule Wat$^{11}$. Hydrogen peroxide is formed and leaves the active site with the assistance of the hydroxide ion product (**7**). Following a rearrangement of the LID loop, Glu$^{47}$ replaces the hydroxide ion (**8**). Finally, the active site is regenerated to its reduced state (**1**) by an unknown external factor. Along the catalytic cycle, *blue* color indicates the (hydro)peroxo species.

**References**

1. M. Costas, M. P. Mehn, M. P. Jensen, L. Que, *Chem. Rev.* **104**, 939 (2004).

2. I. G. Denisov, T. M. Makris, S. G. Sligar, I. Schlichting, *Chem. Rev.* **105**, 2253 (2005).

3. F. E. Jenney, Jr., M. F. Verhagen, X. Cui, M. W. Adams, *Science* **286**, 306 (1999).

4. M. Lombard, D. Touati, M. Fontecave, V. Niviere, *J. Biol. Chem.* **275**, 27021 (2000).

5. G. I. Berglund *et al.*, *Nature* **417**, 463 (2002).

6. I. Schlichting *et al.*, *Science* **287**, 1615 (2000).

7. K. Kuhnel, E. Derat, J. Terner, S. Shaik, I. Schlichting, *Proc. Natl. Acad. Sci. U.S.A.* **104**, 99 (2007).

8. A. Karlsson *et al.*, *Science* **299**, 1039 (2003).

9. D. Bourgeois, A. Royant, *Curr. Opin. Struct. Biol.* **15**, 538 (2005).

10. P. R. Carey, J. Dong, *Biochemistry* **43**, 8885 (2004).

11. L. M. Brines, J. A. Kovacs, *Eur. J. Inorg. Chem.* **2007**, 29 (2007).

12. V. Adam, A. Royant, V. Niviere, F. P. Molina-Heredia, D. Bourgeois, *Structure* **12**, 1729 (2004).

13. A. P. Yeh, Y. Hu, F. E. Jenney, Jr., M. W. Adams, D. C. Rees, *Biochemistry* **39**, 2499 (2000).

14. M. D. Clay *et al.*, *J. Am. Chem. Soc.* **124**, 788 (2002).

15. M. D. Clay *et al.*, *Biochemistry* **45**, 427 (2006).

16. J. P. Emerson, E. D. Coulter, D. E. Cabelli, R. S. Phillips, D. M. Kurtz, Jr., *Biochemistry* **41**, 4348 (2002).

17. V. Niviere *et al.*, *Biochemistry* **43**, 808 (2004).

18. J. V. Rodrigues, I. A. Abreu, D. Cabelli, M. Teixeira, *Biochemistry* **45**, 9266 (2006).

19. C. Mathe *et al.*, *J. Am. Chem. Soc.* **124**, 4966 (2002).

20. C. Mathe, V. Niviere, C. Houee-Levin, T. A. Mattioli, *Biophys. Chem.* **119**, 38 (2006).

21. G. H. Loew, D. L. Harris, *Chem. Rev.* **100**, 407 (2000).

22. C. Mathe, V. Niviere, T. A. Mattioli, *J. Am. Chem. Soc.* **127**, 16436 (2005).

23. G. Roelfes *et al.*, *Inorg. Chem.* **42**, 2639 (2003).

24. Materials and methods are available as supporting material on Science Online.

25. Raman experiments were performed under non-resonant conditions, taking advantage of the large protein concentration found in crystals to enhance the signal to noise ratio. In this way the crystals were sampled homogeneously and potential light-induced damage or photochemistry that could develop under resonant conditions was avoided.

26. J. Girerd, F. Banse, A. J. Simaan, *Struct. Bonding (Berlin)* **97**, 145 (2000).

27. A carefully controlled X-ray exposure was used to collect the diffraction data, so that the dose absorbed by the crystal amounted to ~ 1 % of the Henderson limit, i.e. ~ $2.1 \times 10^5$ Gy.

28. These rearrangements also bring the $N_\varepsilon$ atom of His[119] (the only histidine coordinating the iron through $N_\delta$) to a close distance (3.5 Å) to the distal oxygen of the hydroperoxo moiety.

29. D. Tobi, I. Bahar, *Proc. Natl. Acad. Sci. U.S.A.* **102**, 18908 (2005).

30. M. R. Bukowski, H. L. Halfen, T. A. van den Berg, J. A. Halfen, L. Que, *Angew. Chem. Int. Ed.* **44**, 584 (2005).

31. T. C. Brunold, E. I. Solomon, *J. Am. Chem. Soc.* **121**, 8277 (1999).

**Acknowledgements:**

Coordinates and structure factor amplitudes of the structures: $SOR_{E114A, Fe(II)}$, $SOR_{E114A, Fe(III)}$-OOH and $SOR_{wt, Fe(II)}$ have been deposited in the RCSB Protein Data Bank with pdb codes 2ji2, 2ji3 and 2ji1, respectively. The ESRF is greatly acknowledged for continuous support of our methodological developments. We are grateful to the ESRF beamline staff at ID14-1, ID14-2, ID14-4 and ID29 for their expert assistance. We thank Tony Mattioli and Christelle Mathé for insightful discussions and for providing us with resonance Raman spectra on the E114A SOR mutant. We acknowledge Carole Mathevon for help in sample preparation, Martin Field and Laurent David for help with computational studies, and Hélène Jouve and Jean-Pierre Mahi for providing us with $H_2^{18}O_2$. G.K. acknowledges support by an EMBO long-term fellowship, and D.B. acknowledges support by an "ACI Jeune Chercheurs" from the French Ministry of Research. V. N. acknowledges support by the "Toxicologie Nucléaire" program from the CEA.